Vadim Tarin

# A Polynomial-time Algorithm for Computing the Permanent in GF($3^q$)

**Abstract:**

*A polynomial-time algorithm for computing the permanent in any field of characteristic 3 is presented in this article. The principal objects utilized for that purpose are the Cauchy and Vandermonde matrices, the discriminant function and their generalizations of various types. Classic theorems on the permanent such as the Binet-Minc identity and Borchadt's formula are widely applied, while a special new technique involving the notion of limit re-defined for fields of finite characteristics and corresponding computational methods was developed in order to deal with a number of polynomial-time reductions. All the constructions preserve a strictly algebraic nature ignoring the structure of the basic field, while applying its infinite extensions for calculating limits.*

*A natural corollary of the polynomial-time computability of the permanent in a field of a characteristic different from 2 is the non-uniform equality between the complexity classes P and NP what is equivalent to RP=NP( Ref. [1]).*

**Unless specified otherwise,**

**all the results are for fields of Characteristic 3.**

## Definitions and denotations.

For $n \times m$ -matrices $A, B,$ $A \circ B =^{def} = \{a_{ij} b_{ij}\}_{n \times m}$ , $A^{\bullet h} =^{def} = \{a_{ij}{}^{h}\}_{n \times m}$ ,

$$A^{\bullet(h_1,...,h_q)} =^{def} = \begin{pmatrix} A^{\bullet h_1} \\ ... \\ A^{\bullet h_q} \end{pmatrix}$$ (for real numbers $h, h_1, ..., h_q$)

(the Hadamard product, degree and vector-degree),

for $I \subseteq \{1,...,n\}, J \subseteq \{1,...,m\}$ $A^{(I,J)}$ is the sub-matrix of $A$ whose set of rows is $I$ and of columns $J$ ,

for matrices $A^{[1]},...,A^{[s]}$ with the same number of rows $\coprod_{r=1}^{s} A^{[r]}$ denotes the matrix $\left(A^{[1]} \mid \quad ... \quad \mid A^{[s]}\right)$,

$scal(A) = \sum_{j=1}^{m} \prod_{i=1}^{n} a_{ij}$ ,

$Van^{[h]}(t) =^{def}= \left(t^T\right)^{\bullet(0,...,h-1)}$,

$Van(t) =^{def}= Van^{[\dim(t)]}(t)$ (the Vandermonde matrix),

$Van^{[h]}{}'(t) =^{def}= \frac{d}{dt_1} ... \frac{d}{dt_{\dim(t)}} Van^{[h]}(t)$,

for $\dim(t)=1$ we will write simply $van^{[h]}(t)$, $van^{[h]}{}'(t)$ (as for vectors);

$$W(t) =^{def}= \begin{bmatrix} \begin{pmatrix} Van^{[\dim(t)/2]}(t) \\ Van^{[\dim(t)/2]}(t) \end{pmatrix}, \dim(t) \equiv 0 (\text{mod } 2), \\ \\ \begin{pmatrix} Van^{[(\dim(t)-1)/2]}(t) \\ Van^{[(\dim(t)+1)/2]}(t) \end{pmatrix}, \dim(t) \equiv 1 (\text{mod } 2) \end{bmatrix}$$

$$pol(u,v) =^{def}= \prod_{i=1}^{\dim(u)} \prod_{j=1}^{\dim(v)} (u_i - v_j) ,$$

for a scalar $\tau$, $pol'(\tau,v) =^{def}= \frac{d}{d\tau} pol(\tau,v)$, $pol''(\tau,v) =^{def}= \frac{d^2}{d\tau^2} pol(\tau,v)$ ;

for an appropriate set $I$, $u_I$ is the sub-vector of $u$ consisting of the coordinates with indexes from $I$ (including the case when $I$ is a multi-set, with correspondingly duplicated coordinates), $u_{\setminus I}$ is its compliment in $u$;

for a scalar $\tau$, $\vec{\tau}_{n\times m}$ is the $n \times m$-matrix all whose entries equal $\tau$, for $m=1$ we will write simply $\vec{\tau}_n$ ;` for a set $H$, $Part(H)$ is the set of its partitions on nonempty sets;
for a finite sequence d, |d| denotes the number of its members;

for a real number $u$, $\delta(u) = \begin{bmatrix} 0, u \neq 0 \\ 1, u = 0 \end{bmatrix}$ ;

$C(x,y) =^{def}= \{\frac{1}{x_i - y_j}\}_{\dim(x)\times\dim(y)}$ (the Cauchy matrix);

$$E(a,\phi) =^{def}= \{(1-\delta(i-j))\{\frac{1}{(\varphi_{i,k}!)(\psi_{j,l}!)} \frac{d^{\varphi_{i,k}}}{d\, a_i^{\varphi_{i,k}}} \frac{d^{\psi_{j,l}}}{d\, a_j^{\psi_{j,l}}} \frac{1}{a_i - a_j}\}_{|\varphi_i|\times|\psi_j|}\}_{\dim(a)\times\dim(a)}$$

(the Cauchy extension matrix)

where $\phi = \{\phi_i\}_{\dim(a)} = \{(\varphi_i, \psi_i)\}_{\dim(a)} = \{(\langle\varphi_{i,1},...,\varphi_{i,|\varphi_i|}\rangle, \langle\psi_{i,1},...,\psi_{i,|\psi_i|}\rangle)\}_{\dim(a)}$ is a

$\dim(a)$ -vector whose coordinates are pairs of non-decreasing nonnegative integer finite sequences, including the empty one. The rows and columns containing the term $a_i$ will be called the $a_i$ *-extension* (and $a_i$ itself the extension's *value*); $\phi_i = (\varphi_i, \psi_i)$ we'll call the extension-degree of $a_i$, where $\varphi_i$ is the left extension-degree, $\psi_i$ is the right one, $|\varphi_i|$ is the extension's *height*, $|\psi_i|$ is its *width*, and $bal(\phi_i) =^{def}= |\varphi_i| - |\psi_i|$ is its *balance*. If $|\varphi_i| \cdot |\psi_i| \neq 0$ then we'll say that the matrix has an $a_i$ -singularity. Singularities with $\phi_i = (0,0)$ will be called simple. The sum $bal(\phi) = \sum_{i=1}^{\dim(a)} bal(\phi_i)$ will be called the *total* balance of the extension-degrees' vector $\phi$.

By an E - sum we'll understand the expression

$$\sum_{\phi_1 \in \Omega_1} \cdots \sum_{\phi_{\dim(a)} \in \Omega_{\dim(a)}} (\prod_{i=1}^{\dim(a)} weight_i(\phi_i))\delta(bal(\phi))per(E(a,\phi))$$

where, for $i = 1,...,\dim(a)$, $\Omega_i$ is a set of extension - degrees, $weight_i(\phi_i)$ is its weight - function over the basic field.

The formal sum $\theta_i = \sum_{\phi_i \in \Omega_i} \phi_i \, weight_i(\phi_i)$ will be defined as the extension- *plane* of $a_i$, $weight_i(\phi_i)$ will also be called the *plane-weight* of $\phi_i$ in $\theta_i$, the pair $(a_i, \theta_i)$ will be called an extension-variety and thus we can also talk about the E-sum of $a$ on the extension-planes $\theta_i$, $i = 1,...,\dim(a)$, while the pair of vectors $(a,\theta)$ will be called a vector-variety. If $\theta_i$ has the form $(\varnothing,\varnothing) + \lambda_i(\varnothing,0)$ then we'll say that the E-sum is right-prolonged on $a_i$, if $(\varnothing,\varnothing) + \lambda_i(0,\varnothing)$ then left-prolonged, and the corresponding patterns of $(a,\theta)$ (i. e. sub-vectors of the vector-variety) will be called right and left prolongations (correspondingly).

We will denote the E-sum of $a$ on $\theta$ by $esum(a,\theta)$ (we'll use the brackets $\langle \rangle$ only for left and right extension-degrees whose cardinalities exceed 1);

$$\tilde{C}(x,y,z) =^{def}= E(\begin{pmatrix} x \\ y \\ z \end{pmatrix}, \begin{pmatrix} \overrightarrow{(0,\varnothing)}_{\dim(x)} \\ \overrightarrow{(\varnothing,0)}_{\dim(y)} \\ \overrightarrow{(0,0)}_{\dim(z)} \end{pmatrix}), \quad \tilde{C}(z) =^{def}= E(z, \overrightarrow{(0,0)}_{\dim(z)}) \;;$$

for an $n \times r$ -matrix $A$ and an $m \times r$ -matrix $B$, $m \leq 2n, m$ is even :

$$coper(A,B) =^{def}= \sum_{I,J:\, I \supseteq J,\, |I|=n,\, |J|=m/2} per(A^{(\{1,...,n\},I)})per(B^{(\{1,...,m\},\{J,J\})})$$

(the copermanent of $A$ with $B$),

$$\det(A) = \sum_{J,|J|=n} \det(A^{(\{1,...,n\},J)}) \quad ,$$

$$per(A) = \sum_{J,|J|=n} per(A^{(\{1,...,n\},J)}) \quad ,$$

$$p\hat{e}r_B(A) =^{def}= \sum_{J,|J|=n} per(A^{(\{1,...,n\},J)}) \prod_{k=1}^{m} \sum_{j\in J} b_{kj}$$

(the permanent of $A$ on the base $B$, or the base-permanent of $A$ on $B$);

for r-vectors $u,\alpha,\beta$, nonnegative integers $p \geq q$:

$$dis_{p,q}(u,\alpha,\beta) =^{def}= \sum_{I,J: I\supseteq J, |I|=p, |J|=q} \det{}^2(Van(u_I))\det{}^2(Van(u_J))(\prod_{i\in I}\alpha_i)(\prod_{j\in J}\beta_j)$$

(the $p,q$-discriminantal of $u$ on $\alpha,\beta$) ;

the E-generated functions (for vectors with appropriate dimensions):

the star-function:

$$gen_*(z,u,\rho,\omega^{[1]},\omega^{[2]}) =^{def}=$$
$$=^{def}= \sum_{I,J,\, I\cap J=\varnothing,\, |I\cup J|=\dim(z)/2} (\prod_{i\in I}\omega_i^{[1]})(\prod_{j\in J}\omega_j^{[2]}) per(C(z_{I\cup J}, z_{\backslash I\cup J}))per(C(z_{\{I,J,J\}},u)Diag(\rho))$$

the 2-waves-function:

$$gen_\approx(z,\alpha,w,h,t,\rho) =^{def}=$$
$$=^{def}= \sum_I \sum_J esum(\begin{pmatrix} h \\ t \\ z_I \\ w_J \end{pmatrix}, \begin{pmatrix} \overrightarrow{(0,\varnothing)}_{\dim(x)} \\ \overrightarrow{(\varnothing,\varnothing)}_{\dim(y)} + (\varnothing,0)\rho \\ \overrightarrow{(0,0)}_{|I|} \\ \overrightarrow{\Theta}_{|J|} \end{pmatrix}) esum(\begin{pmatrix} z_{\backslash I} \\ w_{\backslash J} \end{pmatrix}, \begin{pmatrix} (0,0)\alpha_{\backslash I} \\ -\overrightarrow{\Theta}_{\dim(w)-|J|} \end{pmatrix})$$

where $\Theta =^{def}= (\langle 0,0\rangle,\langle 0,0\rangle) + (0,1) - (1,0)$ .

We'll call the extension-plane (0,0) the wave and $\Theta$ the biwave; together with their values (say $z_i$ and $w_j$) such extension-varieties will be called a $z_i$-wave and a $w_j$-biwave (correspondingly) ;

the wave-function (a partial case of the 2-waves-function):

$$gen_\sim(z,\alpha,h,t,\rho) =^{def}= gen_\approx(z,\alpha,\varnothing,h,t,\rho) = \sum_I \sum_{J,|J|=\dim(h)} per(\tilde{C}(h,t_J,z_{\backslash I}))(\prod_{i\in I}\alpha_i)per(\tilde{C}(z_I))\prod_{j\in J}\rho_j \;;$$

For both the wave-function and the 2-waves-function $\alpha_i$ will be called the $i$-th ebb, and $\alpha$ the ebb vector.

the base-function:

$$gen_{\wedge}(u,h,t,\rho) =^{def}= p\hat{e}r_{(-C(u,h)\quad C(u,t))}(\begin{pmatrix} I_{\dim(h)} & \vec{0}_{\dim(h)\times\dim(t)} \\ \vec{0}_{\dim(h)\times\dim(h)} & C(h,t)Diag(\rho) \end{pmatrix}) =$$

$$= \sum_{I}(\prod_{i\in I}(-C(u_i,h)\vec{1}_{\dim(h)}))\, p\hat{e}r_{C(u_{\setminus I},t)}(C(h,t)Diag(\rho))$$

if $F$ is a basic field and $\varepsilon$ is a formal variable then for

$$g = \sum_{k=q}^{\infty} g^{[k]}\varepsilon^k \in F(\varepsilon):\ order_{\varepsilon}(g) = q,\ \lim_{\varepsilon\to 0}(g) = \begin{cases} g^{[0]}, order_{\varepsilon}(g) \geq 0 \\ \text{doesn't exist}, order_{\varepsilon}(g) < 0 \end{cases}$$

By $\Im(f(x),x)$ we'll denote the Jacobian matrix $\{\frac{\partial f_i(x)}{\partial x_j}\}_{\dim(f)\times\dim(x)}$ .

## Preliminaries.

### The Binet-Minc identity (for any characteristic, Ref. [3], [4]):

Let A be an $n\times m-$matrix, then

$$per(A) = (-1)^n \sum_{\mathrm{P}\in Part(\{1,...,n\})} \prod_{I\in \mathrm{P}} \left(-(|I|-1)!\sum_{j=1}^{m}\prod_{i\in I} a_{ij}\right),$$

### The Binet-Minc identity for Characteristic 3:

Let A be an $n\times r-$matrix, then

$$per(A) = (-1)^n \sum_{\mathrm{P}\in Part_{1,2,3}(\{1,...,n\})} \prod_{I\in \mathrm{P}} \left(-(|I|-1)!scal(A^{(I,R)})\right) ,$$

where:

$Part_{1,2,3}(H)$ is the set of partitions of the set $H$ on nonempty sets of cardinalities not bigger than 3, $R = \{1,...,r\}$.

For our purposes we'll also use this identity in the form

$$per(A) = \sum_{\mathrm{P}\in Part_{1,2,3}(\{1,...,n\})} \prod_{I\in \mathrm{P}} \left((-1)^{1+|I|}(|I|-1)!scal(A^{(I,R)})\right)$$

### The generalized Binet-Minc identity for $coper(A,B)$ :

Let $A, B$ be an $n\times r$- and $m\times r$-matrix, $m\equiv 0 (\mathrm{mod}\ 2)$, then

$$coper(A,B) =$$

$$= (-1)^{m/2}\sum_{K\subseteq N}\ \sum_{\substack{\hat{\mathrm{P}}\in Part_{1,2;2}(K,M),\\ \mathrm{P}\in Part_{1,2,3}(N\setminus K)}} (-1)^{n-|K|}\Big(\prod_{(I',I'')\in\hat{\mathrm{P}}} scal(\begin{pmatrix} A^{(I',R)} \\ B^{(I'',R)} \end{pmatrix})\Big)\prod_{I\in\mathrm{P}}\Big((-(|I|-1)!)scal(A^{(I,R)})\Big)$$

where:
$N=\{1,...,n\},\ M=\{1,...,m\},\ R=\{1,...,r\}$,
for two disjoint sets $H'$ and $H''$, $Part_{1,2;2}(H',H'')$ is the set of partitions of the set $H'\cup H''$ on sets having 1 or 2 elements in $H'$ and 2 elements in $H''$ (those sets are denoted here as pairs $(I',I''),\ I'\subseteq H',\ I''\subseteq H''$).

Proof:

the right part can be re-written as

$$\sum_{K\subseteq N}\ \sum_{\substack{\hat{P}\in Part_{1,2;2}(K,M),\\ P\in Part_{1,2,3}(N\backslash K)}}\Big(\prod_{(I',I'')\in\hat{P}} scal(\begin{pmatrix}A^{(I',R)}\\ B^{(I'',R)}\end{pmatrix})\Big)\prod_{I\in P}\Big((|I|-1)!(-1)^{|I|+1}scal(A^{(I,R)})\Big).$$

Let's denote it by $\eta(A,B)$.

Then, by the construction,

$$\eta(A,B)=\sum_{J\subseteq\{1,...,r\}}\ \sum_{\substack{I'_1,I''_1\\ \{I'_1,...,I'_{|J|}\}\in Part(N)\\ \{I''_1,...,I''_{|J|}\}_{\backslash\varnothing}\in Part(M)}}\cdots\sum_{I'_{|J|},I''_{|J|}}\ \prod_{j\in J}\eta(\vec{1}_{|I'_j|},\vec{1}_{|I''_j|})\prod_{i'\in I'_j}\prod_{i''\in I''_j}a_{i'j}b_{i''j}$$

where $\{\ \}_{\backslash\varnothing}$ means the set of sets with removed $\varnothing$-s.
Then let's prove that $\eta(\vec{1}_q,\vec{1}_s)=\delta(q-1)(\delta(s)+\delta(s-2))$ .

## The generalized Binet-Minc identity for $p\hat{e}r_B(A)$:

Let $A$, $B$ be an $n\times r$- and $m\times r$-matrix, then

$$p\hat{e}r_B(A)=\sum_{K\subseteq N}\ \sum_{\substack{\hat{P}\in Part_{1,2;\geq1}(K,M),\\ P\in Part_{1,2,3}(N\backslash K)}}(-1)^{n-|K|}\Big(\prod_{(I',I'')\in\hat{P}}\big((-1)^{\delta(|I'|-2)(1+|I''|)}scal(\begin{pmatrix}A^{(I',R)}\\ B^{(I'',R)}\end{pmatrix})\big)\Big)\prod_{I\in P}\Big((-(|I|-1)!)scal(A^{(I,R)})\Big)$$

where:

$N=\{1,...,n\},\ M=\{1,...,m\},\ R=\{1,...,r\}$,
for two disjoint sets $H'$ and $H''$, $Part_{1,2;\geq1}(H',H'')$ is the set of partitions of the set $H'\cup H''$ on sets having 1 or 2 elements in $H'$ and at least 1 element in $H''$ (those sets are denoted here as pairs $(I',I''),\ I'\subseteq H',\ I''\subseteq H''$).

Proof:

by analogy with the previous proof, we re-write the right part as

$$\sum_{K\subseteq N}\ \sum_{\substack{\hat{P}\in Part_{1,2;\geq1}(K,M),\\ P\in Part_{1,2,3}(N\backslash K)}}\Big(\prod_{(I',I'')\in\hat{P}}\big((-1)^{\delta(|I'|-2)(1+|I''|)}scal(\begin{pmatrix}A^{(I',R)}\\ B^{(I'',R)}\end{pmatrix})\big)\Big)\prod_{I\in P}\Big((|I|-1)!(-1)^{|I|+1}scal(A^{(I,R)})\Big)$$

Let's denote it by $\hat{\eta}(A,B)$.

Then, by the construction,

$$\hat{\eta}(A,B)=\sum_{J\subseteq\{1,\dots,r\}}\ \sum_{\substack{I_1',I_1''}}\cdots\sum_{\substack{I_{|J|}',I_{|J|}'' \\ \{I_1',\dots,I_{|J|}'\}\in Part(N) \\ \{I_1'',\dots,I_{|J|}''\}_{\setminus\varnothing}\in Part(M)}}\ \prod_{j\in J}\hat{\eta}(\vec{1}_{|I_j'|},\vec{1}_{|I_j''|})\prod_{i'\in I_j'}\prod_{i''\in I_j''}a_{i'j}b_{i''j}$$

Then let's prove that $\hat{\eta}(\vec{1}_q,\vec{1}_s)=\delta(q-1)$

## Lemma 1 (known, for any characteristic, Ref. [2], [6]):

Let $\dim(x)=\dim(y)$. Then

$$\det(C(x,y))=(-1)^{\frac{\dim(x)(\dim(x)-1)}{2}}\frac{\det(Van(x))\det(Van(y))}{pol(x,y)}$$

## Corollary 1.1 :

Let $2\dim(x)=\dim(y)$. Then

$$\det(\begin{pmatrix}-C(x,y)\\ C^{\bullet 2}(x,y)\end{pmatrix})=\frac{\det^4(Van(x))\det(Van(y))}{pol^2(x,y)}$$

Proof:

$$\det(\begin{pmatrix}-C(x,y)\\ C^{\bullet 2}(x,y)\end{pmatrix})=\lim_{\varepsilon\to 0}\Big(\varepsilon^{-\dim(x)}\det(\begin{pmatrix}-C(x,y)\\ C(x-\varepsilon\vec{1}_{\dim(x)},y)\end{pmatrix}))\Big)=$$

$$=\lim_{\varepsilon\to 0}\Big(\varepsilon^{-\dim(x)}(-1)^{\dim(x)}\det(C(\begin{pmatrix}x\\ x-\varepsilon\vec{1}_{\dim(x)}\end{pmatrix},y))\Big)$$

## Lemma 2 (for any characteristic, Ref. [5]):

Let $A,B$ be $n\times m$-matrices, $n\le m$. Then

(*)

For $n\equiv 0\ (\text{mod}\ 2)$:

$$\det^2(A)=\det(A\{sign(i-j)\}_{m\times m}A^T),$$

$$\det(A)=\frac{Pf\,(A\{sign(i-j)\}_{m\times m}A^T)}{Pf\,(\{sign(i-j)\}_{n\times n})}$$

where for a real $u$ $sign(u)=\begin{cases}-1, u<0\\ 1, u>0\\ 0, u=0\end{cases}$

For $\forall n,\ q = 1,...,m,$ scalar $\tau$ : $\det(A \begin{pmatrix} I_q & \vec{0}_{q\times 2} & \vec{0}_{q\times(m-q-2)} \\ \vec{0}_{2\times q} & \begin{pmatrix} \tau & 1-\tau \\ \tau-1 & 2-\tau \end{pmatrix} & \vec{0}_{2\times(m-q-2)} \\ \vec{0}_{(m-q-2)\times q} & \vec{0}_{(m-q-2)\times 2} & I_{m-q-2} \end{pmatrix}) = \det(A)$.

In the generic case this method allows to reduce $A$ to a triangular matrix when computing $\det(A)$.

(**) for $n \equiv 0 \pmod 2$

$$\sum_{J,|J|=n/2} \det(\coprod_{j\in J} \left(A^{(N,j)} \quad | \quad B^{(N,j)}\right)) = \det(\coprod_{k=1}^{m} \left(A^{(N,k)} \quad | \quad B^{(N,k)} \quad | \quad -A^{(N,k)} - B^{(N,k)}\right)) \ ,$$

where $N = \{1,...,n\}$.

## Lemma 3 (for any characteristic):

Let $f(x)$ be a polynomial in $x_1,...,x_{\dim(x)}$ of degree $d$. Then $f(\chi) = \sum_{i=0}^{d} coef_{\varepsilon^i}(f(x + \varepsilon(-x+\chi)))$ where $coef_{\varepsilon^i}$ is the coefficient at $\varepsilon^i$ (if the expression is considered as a polynomial in $\varepsilon$).

## The neighbouring computation principle (for any characteristic):

Let $f(u)$ be a polynomial in $u_1,...,u_{\dim(u)}$ of degree $d$ over the field $F$,

$u = u(h_1,...,h_{\dim(h)})$ , $u$ exists at $h^{[0]}$;

$v = v(h_1,...,h_{\dim(h)})$, $v(h^{[0]}) = \vec{0}_{\dim(v)}$;

$\Im(\begin{pmatrix} u \\ v \end{pmatrix}, h^{[0]})$ exists and is nonsingular.

Then, given $\upsilon$, over $F(\varepsilon)$:

$$f(\upsilon) = \sum_{i=0}^{d} coef_{\varepsilon^i}(f(u(\sum_{k=0}^{d} \varepsilon^k h^{[k]}))) .$$

where:

$$\Im(\begin{pmatrix} u \\ v \end{pmatrix}, h^{[0]}) h^{[k]} = \begin{pmatrix} \delta(k-1)(u(h^{[0]}) - \upsilon) - coef_{\varepsilon^k}(u(\sum_{i=0}^{k-1} \varepsilon^i h^{[i]})) \\ \vec{0}_{\dim(v)} \end{pmatrix} ,\ k = 1,...,d$$

(And thus $u(\sum_{k=0}^{d} \varepsilon^k h^{[k]}) = u(h^{[0]}) + \varepsilon(-u(h^{[0]}) + \upsilon) + O(\varepsilon^{d+1})$ )

This method will be called the *neighbouring* computation of the function $f(\upsilon)$ by the *parameterization* $u(h)$ in the *region* $v(h) = \vec{0}_{\dim(v)}$ from the *bearing* point $h^{[0]}$.

If $f(u(h))$ is computable in polynomial time for any $h$ in the region $v(h) = \vec{0}_{\dim(v)}$ and there exists a bearing point, then $f(\upsilon)$ is computable in polynomial time for any $\upsilon$ too.

In the further, to be clear and short, we'll call a system of functions S algebraically *absolutely* independent in a region R (given by a system of equations with a zero right part) iff the joint system of functions consisting of S and the left part of the system representing R is algebraically independent at some point of R.

## The prolongation-derivative principle (for any characteristic):

a left (right) prolongation pattern of a vector-variety can be removed with an induced coordinate-wise transformation of the remaining vector of extension-planes such that the E-sum preserves its value. Such a transformation will be called a left (right) prolongation-derivative, or the prolongation-derivative on the given prolongation. Formally it will be denoted (correspondingly for the left and right cases) by

$$\frac{d_L}{d_L(\gamma,\lambda)} esum(\beta,\vartheta) =^{def}= esum(\beta,\frac{d_L}{d_L(\beta,\gamma,\lambda)}\vartheta) =^{def}= esum(\begin{pmatrix}\beta\\ \gamma\end{pmatrix},\begin{pmatrix}\vartheta\\ \overrightarrow{(\varnothing,\varnothing)}_{\dim(\gamma)} + (1,\varnothing)\lambda\end{pmatrix})$$

$$\frac{d_R}{d_R(\gamma,\lambda)} esum(\beta,\vartheta) =^{def}= esum(\beta,\frac{d_R}{d_R(\beta,\gamma,\lambda)}\vartheta) =^{def}= esum(\begin{pmatrix}\beta\\ \gamma\end{pmatrix},\begin{pmatrix}\vartheta\\ \overrightarrow{(\varnothing,\varnothing)}_{\dim(\gamma)} + (\varnothing,1)\lambda\end{pmatrix})$$

This principle is based on the Binet-Minc identity.

## Lemma 4:

Let A be an $n \times n$-matrix, B be a $(2n) \times n$-matrix. Then

$$per((BA \quad | \quad BA)) = per((B \quad | \quad B))\det^2(A)$$

## The Borchardt formula (for any characteristic, Ref. [2]) :

Let $\dim(x) = \dim(y)$. Then

$$per(C(x,y)) = \frac{\det(C^{\bullet 2}(x,y))}{\det(C(x,y))}$$

## Lemma 5:

$$per(W(t)) = \frac{\det(t^{\bullet \eta_{\dim(t)}})}{\det(Van(t))}$$

where $\eta_{\dim(t)}$ is the first $\dim(t)$ members of the sequence composed as the sequence of pairs $(3s, 3s+1),\ s = 0,...,\infty$

Proof:

$$per(W(t)) = \lim_{\varepsilon \to 0} \frac{per(C(\varepsilon \cdot x, t))}{per(W(\varepsilon \cdot x)) \prod_{j=1}^{\dim(y)} t_j}$$

# The main part.

## Theorem I :

Let $\dim(t) = 2\dim(u) = 2n$.

Then

$$per(C(t, \begin{pmatrix} u \\ u \end{pmatrix})) = per(W(t)) \frac{\det^2(Van(u))}{pol(t,u)}$$

Proof:

$$per(C(t,u)) = \frac{per(Diag(\{pol(t_i,u)\}_{2n})C(t,u))}{\prod_{i=1}^{2n} pol(t_i,u)} =$$

$$= \frac{per(W^T(t)\begin{pmatrix} R(u) & 0 \\ 0 & R(u) \end{pmatrix})}{pol(t,u)} = \frac{per(W^T(t))}{pol(t,u)} \det^2(R(u))$$

where $R(u)$ is the matrix of the coefficients of the polynomials in the formal scalar variable $\tau$ $pol(\tau, u_{\backslash 1}), ..., pol(\tau, u_{\backslash n})$. The determinant of this matrix is a homogenous polynomial in $u_1, ..., u_n$ of degree $0 + 1 + ... + (n-1) = \frac{n(n-1)}{2}$, and in the meantime it is divided by $u_i - u_j$ for any $1 \le i < j \le n$, or by $\det(Van(u))$; moreover, its coefficients are 1 and -1, therefore $\det^2(R(u)) = \det^2(Van(u))$, what completes the proof.

## Corollary I.1 :

Let $\dim(t) = 2\dim(u) + \dim(v)$ .Then

$$per(C(t, \begin{pmatrix} u \\ u \\ v \end{pmatrix})) = per(W(\begin{pmatrix} t \\ v \end{pmatrix})) \frac{\det^2(Van(u)) pol(u,v)}{pol(t, \begin{pmatrix} u \\ v \end{pmatrix})}$$

Proof:

$$per(C(t,\begin{pmatrix}u\\u\\v\end{pmatrix}))=\lim_{\varepsilon\to 0}per(\varepsilon^{\dim(v)}C(\begin{pmatrix}t\\v-\varepsilon\vec{1}_{\dim(v)}\end{pmatrix},\begin{pmatrix}u\\u\\v\\v\end{pmatrix}))$$

and, basing on Theorem I, we calculate the limit to get the lemma's equality.

## Corollary I.2 :

Let $\dim(t)=2\dim(u)+\dim(v)+1$. Then

$$per(C(t,\begin{pmatrix}u\\u\\v\end{pmatrix})\quad \vec{1}_{\dim(t)})=per(W(\begin{pmatrix}t\\v\end{pmatrix}))\frac{\det^2(Van(u))pol(u,v)}{pol(t,\begin{pmatrix}u\\v\end{pmatrix})}$$

Proof:

$$per(C(t,\begin{pmatrix}u\\u\\v\end{pmatrix})\quad \vec{1}_{\dim(t)})=\lim_{\varepsilon\to 0}(per(C(t,\begin{pmatrix}u\\u\\v\\1/\varepsilon\end{pmatrix}))/\varepsilon)$$

and then, basing on Corollary I.1, we again just calculate the limit.

## Theorem II :

$$dis_{p,q}(u,\alpha,\beta)=\det(\coprod_{i=1}^{\dim(u)}\begin{pmatrix}\alpha_i van^{[p+q]}(u_i) & \beta_i van^{[p+q]}{}'(u_i) & -\alpha_i van^{[p+q]}(u_i)-\beta_i van^{[p+q]}{}'(u_i)\\ \vec{0}_{p-q} & van^{[p-q]}(u_i) & -van^{[p-q]}(u_i)\end{pmatrix})$$

Proof:

basing on Lemma 2, we conclude that the right part of this equality is

$$\sum_{I,|I|=p}\det(\coprod_{i\in I}\begin{pmatrix}\alpha_i van^{[p+q]}(u_i) & \beta_i van^{[p+q]}{}'(u_i)\\ \vec{0}_{p-q} & van^{[p-q]}(u_i)\end{pmatrix})$$

## Theorem III :

Let

$|I|=\dim(x)=n,\dim(y)=2n,\dim(t)=3n+1, pol''(\tau,\begin{pmatrix}x\\y\end{pmatrix})\equiv 0,$

$pol'(\tau,t)\equiv pol(\tau,\begin{pmatrix}x\\y\end{pmatrix})$

Then

$$per(C(x,y_I)) = \frac{per(W(\begin{pmatrix} t \\ x \end{pmatrix}))}{pol(x,t)} \det^2(Van(y_I)) \prod_{i \in I} \frac{1}{pol'(y_i,y)}$$

Proof:

Let us apply the Binet-Minc identity to $per((C(t,\begin{pmatrix} x \\ y_I \\ y_I \end{pmatrix}) \ | \ \vec{1}_{\dim(t)})^T)$

### Theorem IV :

Let

$$\dim(x) = n, \dim(y) = 2n, \dim(t) = 3n+1, \dim(z) = m \equiv 0 \ (\mathrm{mod}\ 2)$$

$$pol''(\tau,\begin{pmatrix} x \\ y \end{pmatrix}) \equiv 0, pol'(\tau,t) \equiv pol(\tau,\begin{pmatrix} x \\ y \end{pmatrix})$$

Then

$$coper(C(x,y)Diag(\alpha), C(z,y)Diag(\beta)) =$$

$$= \frac{per(W(\begin{pmatrix} t \\ x \end{pmatrix}))\, per(W(z))}{pol(x,t)} dis_{n,m/2}(y, \{\frac{\alpha_i}{pol'(y_i,y)}\}_{2n}, \{\frac{\beta_i^2}{pol(y_i,z)}\}_{2n})$$

Proof:

It follows from the definition of the copermanent and Theorems I, III.

-----------------------------------------------------------------------------------------

Let's note that for $pol''(\tau,\begin{pmatrix} x \\ y \end{pmatrix}) \equiv 0$ to be true it's sufficient that

$pol''(\tau,\begin{pmatrix} x \\ y \end{pmatrix}) = 0$ for any $n$ pair-wise distinct values of $\tau$ (due to the characteristic),

for instance for all the coordinates of $x$, i.e. it's equivalent to $\{pol''(x_i,\begin{pmatrix} x \\ y \end{pmatrix})\}_n = \vec{0}_n$ ,

or $\{pol''(x_i,\begin{pmatrix} x \\ y \end{pmatrix}) / pol'(x_i,\begin{pmatrix} x \\ y \end{pmatrix})\}_n = \tilde{C}(x)\vec{1}_n + C(x,y)\vec{1}_{2n} = \vec{0}_n$ .

### Theorem V :

Let

$$\dim(x) = n, \dim(z) = m \equiv 0 \ (\mathrm{mod}\ 2),$$

$$C(x,y)\lambda^{\bullet h} = \mu_h \vec{1}_n, \quad h = 2,3 .$$

Then

$$gen_*(z,x,(C(x,y)\lambda)^{\bullet(-1)},C(z,y)\lambda^{\bullet 2}-\mu_2\vec{1}_{\dim(z)},C(z,y)\lambda^{\bullet 3}-\mu_3\vec{1}_{\dim(z)})=$$

$$=(-1)^{m/2}\frac{coper(C(x,y)Diag(\lambda),C(z,y)Diag(\lambda^{\bullet(1/2)}))}{\prod_{s=1}^{n}C(x_s,y)\lambda}$$

Proof:

It follows from the Binet-Minc identity for $coper(A,B)$

## Theorem VI :

Let $\dim(x)=n, \dim(y)=2n, \dim(z)=m<<n$.

Then there exist scalar constants $\mu_2,\mu_3$ such that the system of functions in $x,y,\lambda$

$$(\$)\quad k=1,..,m:\begin{cases}C(z_k,y)\lambda^{\bullet s} & ,s=2,3\\ C^{\bullet q}(z_k,x)(C(x,y)\lambda)^{\bullet(-s)} & ,q=1,2,\ s=q,...,3\end{cases}$$

is algebraically absolutely independent in the region

$$(\Re)\begin{cases}\tilde{C}(x)\vec{1}_n+C(x,y)\vec{1}_{2n}=\vec{0}_n & ,\\ C(x,y)\lambda^{\bullet s}-\mu_s\vec{1}_n=\vec{0}_n & ,\quad s=2,3\end{cases}$$

Proof:

let us consider the Jacobian matrix which is

$$\Im\left(\begin{pmatrix}C(\begin{pmatrix}x\\z\end{pmatrix},y)\lambda^{\bullet 2}\\ C(\begin{pmatrix}x\\z\end{pmatrix},y)\lambda^{\bullet 3}\\ C(z,x)(C(x,y)\lambda)^{\bullet(-1)}\\ C^{\bullet(1,2)}(z,x)(C(x,y)\lambda)^{\bullet(-2)}\\ C^{\bullet(1,2)}(z,x)(C(x,y)\lambda)^{\bullet(-3)}\\ \{pol''(x_i,\begin{pmatrix}x\\y\end{pmatrix})\}_n\end{pmatrix},\begin{pmatrix}x\\y\\\lambda\end{pmatrix}\right)=$$

$$= \begin{pmatrix} * & * & -C(\begin{pmatrix} x \\ z \end{pmatrix}, y)Diag(\lambda) \\ \begin{pmatrix} -Diag(C^{\bullet 2}(x,y)\lambda^{\bullet 3}) \\ \vec{0}_{m\times(2n)} \end{pmatrix} & C^{\bullet 2}(\begin{pmatrix} x \\ z \end{pmatrix}, y)Diag(\lambda^{\bullet 3}) & \vec{0}_{(m+n)\times(2n)} \\ * & * & -C(z,x)Diag((C(x,y)\lambda)^{\bullet(-2)})C(x,y) \\ * & * & C^{\bullet(1,2)}(z,x)Diag((C(x,y)\lambda)^{\bullet(-3)})C(x,y) \\ \begin{pmatrix} -C^{\bullet 2}(z,x)Diag((C(x,y)\lambda)^{\bullet(-3)}) \\ C^{\bullet 3}(z,x)Diag((C(x,y)\lambda)^{\bullet(-3)}) \end{pmatrix} & \vec{0}_{(2m)\times(2n)} & \vec{0}_{(2m)\times(2n)} \\ \tilde{C}^{\bullet 2}(x) - Diag(\tilde{C}^{\bullet 2}(x)\vec{1}_n + C(x,y)\vec{1}_{2n}) & C^{\bullet 2}(x,y) & \vec{0}_{n\times(2n)} \end{pmatrix}$$

where $*$-s denote some matrices of appropriate dimensions. For the purpose of simplicity, let's permute in a certain way the Jacobian's rows and multiply some of them by -1, thus reducing it to the form $\begin{pmatrix} A_{11} & \vec{0}_{(3m+2n)\times(2n)} \\ A_{21} & A_{22} \end{pmatrix}$ where

$$A_{11} = \begin{pmatrix} -Diag(C^{\bullet 2}(x,y)\lambda^{\bullet 3}) & C^{\bullet 2}(x,y)Diag(\lambda^{\bullet 3}) \\ \tilde{C}^{\bullet 2}(x) - Diag(\tilde{C}^{\bullet 2}(x)\vec{1}_n + C^{\bullet 2}(x,y)\vec{1}_{2n}) & C^{\bullet 2}(x,y) \\ C^{\bullet(2,3)}(z,x)Diag((C(x,y)\lambda)^{\bullet(-3)}) & \vec{0}_{(2m)\times(2n)} \\ \vec{0}_{m\times(2n)} & C^{\bullet 2}(z,y)Diag(\lambda^{\bullet 3}) \end{pmatrix}$$

$$A_{22} = \begin{pmatrix} -C(\begin{pmatrix} x \\ z \end{pmatrix}, y)Diag(\lambda) \\ -C(z,x)Diag((C(x,y)\lambda)^{\bullet(-2)})C(x,y) \\ C^{\bullet(1,2)}(z,x)Diag((C(x,y)\lambda)^{\bullet(-3)})C(x,y) \\ \end{pmatrix}$$

If both the matrices are nonsingular then the Jacobian matrix is nonsingular too. If $F$ is the basic field, let's consider them over $F(\dot{\varepsilon})$ for

$x = x^{[0]} + x^{[1]}\dot{\varepsilon},\ \ x^{[0]} = h \otimes \vec{1}_{\dim(x)/\dim(h)}$,

$y = \sum_{k=0}^{\infty} y^{[k]}\dot{\varepsilon}^k,\ \ y^{[0]} = h \otimes \vec{1}_{2\dim(x)/\dim(h)}$ ,

$\lambda = \sum_{k=0}^{\infty} \lambda^{[k]}\dot{\varepsilon}^k$,

where $h, x^{[1]}, y^{[k]}, \lambda^{[k]}\ (\forall k)$ are vectors over $F$, $\dim(x) \equiv 0(\mathrm{mod}\,\dim(\dot{x}))$.

First we'll show how an $\Re$-point (satisfying the region's conditions $(\Re)$) of such a type is built and then we'll prove that in the generic case it gives nonsingular $A_{11}, A_{22}$.

In order to build such an $\Re$-point, it's sufficient that the Jacobian matrix of the system of functions representing the left part of $(\Re)$ exists and is nonsingular at a certain approximation of the power series (in $\dot{\varepsilon}$) for $y, \lambda$ which satisfies the region's conditions up to some degree of $\dot{\varepsilon}$ (then all the other members of the power series are computable via corresponding linear equations involving the Jacobian matrix calculated for the approximation). Let's consider only the first two members of the power series for $y$ (i.e. $y^{[0]} + y^{[1]}\dot{\varepsilon}$) and the first member of the one for $\lambda$ (i.e. $\lambda^{[0]}$).

For $A_{11}$, let's multiply its first two block-rows by $\dot{\varepsilon}^2$ and then consider its $\lim_{\dot{\varepsilon}\to 0}$. In this limit let's substitute the matrix received via summing up all the columns corresponding to the same $h_r$ in $y^{[0]}, r = 1,...,\dim(h)$, for its first block-column (such a transformation can't enlarge the rank) .

And now $A_{22}$:

$$A_{22} = \begin{pmatrix} -C(\begin{pmatrix} x \\ z \end{pmatrix}, y)Diag(\lambda) \\ -C(z,x)Diag((C(x,y)\lambda)^{\bullet(-2)})C(x,y) \\ C^{\bullet(1,2)}(z,x)Diag((C(x,y)\lambda)^{\bullet(-3)})C(x,y) \end{pmatrix} = \begin{pmatrix} -C(x,y)Diag(\lambda) \\ -C(z,y)Diag(\lambda) \\ -C(z,x)Diag((C(x,y)\lambda)^{\bullet(-2)})C(x,y) \\ C^{\bullet(1,2)}(z,x)Diag((C(x,y)\lambda)^{\bullet(-3)})C(x,y) \end{pmatrix} =$$

$$= \begin{pmatrix} -C(x,y)Diag(\lambda) \\ -\{\frac{(-1)^{n+j}}{\det(P)}\det\begin{pmatrix} C(x,y)Diag(\lambda) \\ C(x_{\setminus j},y) \\ C(z_i,y)Diag(\lambda) \end{pmatrix}\}_{m\times n}C(x,y) \\ -C(z,x)Diag((C(x,y)\lambda)^{\bullet(-2)})C(x,y) \\ C^{\bullet(1,2)}(z,x)Diag((C(x,y)\lambda)^{\bullet(-3)})C(x,y) \end{pmatrix} = \begin{pmatrix} -I_{n\times n} & \vec{0}_{n\times n} \\ \vec{0}_{m\times n} & -\{\frac{(-1)^{n+j}}{\det(P)}\det\begin{pmatrix} C(x,y)Diag(\lambda) \\ C(x_{\setminus j},y) \\ C(z_i,y)Diag(\lambda) \end{pmatrix}\}_{m\times n} \\ \vec{0}_{m\times n} & -C(z,x)Diag((C(x,y)\lambda)^{\bullet(-2)}) \\ \vec{0}_{m\times n} & C^{\bullet(1,2)}(z,x)Diag((C(x,y)\lambda)^{\bullet(-3)}) \end{pmatrix} P$$

where $P = \begin{pmatrix} C(x,y)Diag(\lambda) \\ C(x,y) \end{pmatrix}$ .

Since we use the expression $\frac{1}{\det(P)}$, first let's prove the non-singularity of $P$ in the generic case. It's clearly seen for $\dim(h) = \dim(x)$ .

## Corollary VI.1 :

$gen_*(z,u,\rho,\omega^{[1]},\omega^{[2]})$ (which, according to the Binet-Minc identity, is a function in the *essential* variables: the *Cauchy - Binet - Minc weights* $C^{\bullet q}(z_i,u)\rho^{\bullet s}$ $(q = 1,2,\ s = q,...,3)$ and the *star - functional weights* $\omega_i^{[1]},\omega_i^{[2]}$) is computable in polynomial time for any values of the essential (and, hence, original) variables via a neighbouring computation based on Theorems V, VI as the following:

the parameterization (of the essential variables) in $x, y, \lambda$ :

$$\begin{cases} \dim(z) = m \equiv 0 (\text{mod } 2) \\ \omega^{[s]} = C(z, y)\lambda^{\bullet s+1} - \mu_{s+1}\vec{1}_m \quad , s = 1,2 \\ C^{\bullet q}(z, u)\rho^{\bullet s} = C(z, x)(C(x, y)\lambda)^{\bullet(-s)}, \, q = 1,2, \, s = q,...,3 \end{cases}$$

the region : $\begin{cases} \quad \dim(y) = 2\dim(x) = 2n \\ \quad \tilde{C}(x)\vec{1}_n + C(x, y)\vec{1}_{2n} = \vec{0}_n \quad , \\ C(x, y)\lambda^{\bullet s} - \mu_s\vec{1}_n = \vec{0}_n \quad , \quad s = 2,3 \end{cases}$

Hence the <u>final formula</u> for $gen_*(z, u, \rho, \omega^{[1]}, \omega^{[2]})$ is:

$$gen_*(z, u, \rho, \omega^{[1]}, \omega^{[2]}) = \frac{per(W(\begin{pmatrix} t \\ x \end{pmatrix}))\, per(W(z))}{pol(x, t)} dis_{n, m/2}(y, \{\frac{\lambda_i}{pol'(y_i, y)}\}_{2n}, \{\frac{\lambda_i}{pol(y_i, z)}\}_{2n})$$

where: $\dim(t) = 3n + 1, \dim(y) = 2\dim(x) = 2n, \dim(z) = m \equiv 0 (\text{mod } 2)$ ,

$$pol'(\tau, t) \equiv pol(\tau, \begin{pmatrix} x \\ y \end{pmatrix}) \ ,$$

the region : $\begin{cases} \quad \tilde{C}(x)\vec{1}_n + C(x, y)\vec{1}_{2n} = \vec{0}_n \quad , \\ C(x, y)\lambda^{\bullet s} - \mu_s\vec{1}_n = \vec{0}_n \quad , \quad s = 2,3 \end{cases}$

the parameterization : $\begin{cases} \omega^{[s]} = C(z, y)\lambda^{\bullet(s+1)} - \mu_{s+1}\vec{1}_m \quad , s = 1,2 \\ C^{\bullet q}(z, u)\rho^{\bullet s} = C^{\bullet q}(z, x)(C(x, y)\lambda)^{\bullet(-s)}, \, q = 1,2, \, s = q,...,3 \end{cases}$

( $\mu_2, \mu_3$ are scalar constants)

**Theorem VII :**

$$gen_{\sim}(\begin{pmatrix} u \\ v \end{pmatrix}, \begin{pmatrix} \vec{1}_{\dim(u)} \\ -\vec{1}_{\dim(v)} \end{pmatrix}, h, t, \rho) =$$

$$= \lim_{\varepsilon \to 0}(\varepsilon^{2\dim(u)+\dim(v)+\dim(w)} gen_{*}(\begin{pmatrix} u+\varepsilon\vec{1}_{\dim(u)} \\ u-\varepsilon\vec{1}_{\dim(u)} \\ v+\varepsilon\vec{1}_{\dim(v)} \\ v-\varepsilon\vec{1}_{\dim(v)} \\ h+\varepsilon\vec{1}_{\dim(w)} \\ h \end{pmatrix}, \begin{pmatrix} u \\ v \\ v \\ t \end{pmatrix}, \begin{pmatrix} (1/\varepsilon)\vec{1}_{\dim(u)} \\ \vec{1}_{\dim(v)} \\ -\vec{1}_{\dim(v)} \\ \rho \end{pmatrix}, \begin{pmatrix} \vec{1}_{\dim(u)} \\ -\vec{1}_{\dim(u)} \\ \vec{0}_{\dim(v)} \\ \vec{0}_{\dim(v)} \\ \vec{1}_{\dim(w)} \\ \vec{0}_{\dim(w)} \end{pmatrix}, \begin{pmatrix} \vec{0}_{\dim(u)} \\ \vec{0}_{\dim(u)} \\ \varepsilon\vec{1}_{\dim(v)} \\ \varepsilon\vec{1}_{\dim(v)} \\ \vec{0}_{\dim(w)} \\ \vec{0}_{\dim(w)} \end{pmatrix}))$$

**Theorem VIII :**

$$gen_{\approx}(z, \alpha, w, h, t, \rho) = \lim_{\varepsilon \to 0}(gen_{\sim}(\begin{pmatrix} z \\ w \\ w+\varepsilon\vec{1}_{\dim(u)} \end{pmatrix}, \begin{pmatrix} \alpha \\ \vec{1}_{\dim(u)} \\ -\vec{1}_{\dim(u)} \end{pmatrix}, h, t, \rho))$$

**Theorem IX :**

$$gen_{\sim}(\begin{pmatrix} z \\ w \end{pmatrix}, \begin{pmatrix} \alpha \\ \vec{0}_{\dim(u)} \end{pmatrix}, h, t, \rho) = (\prod_{j=1}^{\dim(x)} \lambda_j) \lim_{\varepsilon \to 0}\Big(\varepsilon^{2\dim(w)} coef_{\xi^{\dim(h)}}(gen_{\approx}(z, \alpha, w, \begin{pmatrix} h \\ x \end{pmatrix}, \begin{pmatrix} t \\ y \end{pmatrix}, \begin{pmatrix} \xi\rho \\ \gamma \end{pmatrix}))\Big)$$

where $\xi$ is a formal scalar variable, and for $s = 1,2,\ q = 1,2,3$:

(i) $$\begin{pmatrix} C^{\bullet s}(w,x) \\ C(\begin{pmatrix} z \\ t \end{pmatrix}, x) \end{pmatrix} \lambda^{\bullet q} = \begin{pmatrix} \overrightarrow{\big((-1)^{1+s}\varepsilon^{-s}\delta(s-q) - \varepsilon^{-3}\delta(s-2)\delta(q-1)\big)}_{\dim(w)} \\ \vec{0}_{\dim(t)+\dim(z)} \end{pmatrix}$$

(ii) $$\begin{pmatrix} (-1)^s C^{\bullet s}(w,y) \\ -C(\begin{pmatrix} z \\ h \end{pmatrix}, y) \\ -C(x,y) \end{pmatrix} \gamma^{\bullet q} = \begin{pmatrix} \overrightarrow{\big(\varepsilon^{-s}\delta(s-q) + \varepsilon^{-3}\delta(s-2)\delta(q-1)\big)}_{\dim(w)} \\ \vec{0}_{\dim(h)+\dim(z)} \\ \delta(q-1)\lambda^{\bullet(-1)} \end{pmatrix}$$

Proof:

According to the definition, $coef_{\xi^{\dim(h)}}(gen_{\approx}(z, \alpha, w, \begin{pmatrix} h \\ x \end{pmatrix}, \begin{pmatrix} t \\ y \end{pmatrix}, \begin{pmatrix} \xi\rho \\ \beta \end{pmatrix})) =$

$$= \sum_{I}\sum_{J}\sum_{K, |K|=\dim(h)} esum(\begin{pmatrix} h \\ x \\ t_K \\ y \\ z_I \\ w_J \end{pmatrix}, \begin{pmatrix} \overrightarrow{(0,\varnothing)}_{\dim(h)} \\ \overrightarrow{(0,\varnothing)}_{\dim(x)} \\ (\varnothing,0)\rho_K \\ \overrightarrow{(\varnothing,\varnothing)}_{\dim(y)} + (\varnothing,0)\gamma \\ \overrightarrow{(0,0)}_{|I|} \\ \vec{\Theta}_{|J|} \end{pmatrix}) esum(\begin{pmatrix} z_{\backslash I} \\ w_{\backslash J} \end{pmatrix}, \begin{pmatrix} (0,0)\alpha_{\backslash I} \\ -\vec{\Theta}_{\dim(w)-|J|} \end{pmatrix})$$

and in such a case $(y, \overrightarrow{(\varnothing,\varnothing)}_{\dim(y)} + (\varnothing,0)\gamma)$ is a right prolongation. Then, according to the prolongation-derivative principle and the equality (ii), the first multiplier of the expression under the summation signs is

$$esum(\begin{pmatrix} h \\ x \\ t_K \\ z_I \\ w_J \end{pmatrix}, \begin{pmatrix} \overrightarrow{(0,\varnothing)}_{\dim(h)} \\ (\varnothing,\varnothing)\lambda^{\bullet(-1)} + \overrightarrow{(0,\varnothing)}_{\dim(x)} \\ (\varnothing,0)\rho_K \\ \overrightarrow{(0,0)}_{|I|} \\ \overrightarrow{\left(\Theta + \varepsilon^{-1}(\varnothing,1) - \varepsilon^{-1}(0,\langle 0,0\rangle) + \varepsilon^{-2}(\varnothing,\langle 0,0\rangle) - \varepsilon^{-3}(\varnothing,0)\right)}_{|J|} \end{pmatrix}) =$$

$$= \frac{1}{\prod_{j=1}^{\dim(x)} \lambda_j} esum(\begin{pmatrix} h \\ x \\ t_K \\ z_I \\ w_J \end{pmatrix}, \begin{pmatrix} \overrightarrow{(0,\varnothing)}_{\dim(h)} \\ \overrightarrow{(\varnothing,\varnothing)}_{\dim(x)} + (0,\varnothing)\lambda \\ (\varnothing,0)\rho_K \\ \overrightarrow{(0,0)}_{|I|} \\ \overrightarrow{\left(\Theta + \varepsilon^{-1}(\varnothing,1) - \varepsilon^{-1}(0,\langle 0,0\rangle) + \varepsilon^{-2}(\varnothing,\langle 0,0\rangle) - \varepsilon^{-3}(\varnothing,0)\right)}_{|J|} \end{pmatrix})$$

Then $(x, \overrightarrow{(\varnothing,\varnothing)}_{\dim(x)} + (0,\varnothing)\lambda)$ is a left prolongation, and we can apply the prolongation-derivative principle again, this time together with the equality (i), hence receiving

$$\frac{1}{\prod_{j=1}^{\dim(x)} \lambda_j} esum(\begin{pmatrix} h \\ t_K \\ z_I \\ w_J \end{pmatrix}, \begin{pmatrix} \overrightarrow{(0,\varnothing)}_{\dim(h)} \\ (\varnothing,0)\rho_K \\ \overrightarrow{(0,0)}_{|I|} \\ \overrightarrow{\left(o(\varepsilon^{-1}) + \varepsilon^{-2}(0,0) - \varepsilon^{-2}(\varnothing,\langle 0,0\rangle)\right)}_{|J|} \end{pmatrix})$$

Because the second multiplier doesn't depend on $\varepsilon$, we can apply the operator $\lim_{\varepsilon\to 0}(\varepsilon^{2\dim(w)}...)$ to the first one only, hence receiving

$$\frac{esum(\begin{pmatrix} h \\ t_K \\ z_I \\ w \end{pmatrix}, \begin{pmatrix} \overrightarrow{(0,\varnothing)}_{\dim(h)} \\ (\varnothing,0)\rho_K \\ \overrightarrow{(0,0)}_{|I|} \\ \overrightarrow{\left((0,0) - (\varnothing,\langle 0,0\rangle)\right)}_{\dim(w)} \end{pmatrix})}{\prod_{j=1}^{\dim(x)} \lambda_j} \delta(|J| - \dim(w)) = \frac{esum(\begin{pmatrix} h \\ t_K \\ z_I \\ w \end{pmatrix}, \begin{pmatrix} \overrightarrow{(0,\varnothing)}_{\dim(h)} \\ (\varnothing,0)\rho_K \\ \overrightarrow{(0,0)}_{|I|} \\ \overrightarrow{(0,0)}_{\dim(w)} \end{pmatrix})}{\prod_{j=1}^{\dim(x)} \lambda_j} \delta(|J| - \dim(w))$$

where, taking into account that the third sum $\sum_{K,|K|=\dim(h)}$ runs over sets of cardinality $\dim(h)$ (thus "balancing" the patterns $\overrightarrow{(0,\varnothing)}_{\dim(h)}$ and $(\varnothing,0)\rho_K$), the last passage is due to the requirement (following from the E-sum's definition) that the total balance is to be 0. So, after subjecting the product of the two multipliers to the overall summation $\sum_I \sum_J \sum_{K,|K|=\dim(h)}$ , we get the theorem's claim.

## Theorem X (for any characteristic) :

1) $per(\tilde{C}(z)) = \sum_{P \in Part_2(\{1,...,\dim(z)\})} \prod_{\{i,j\} \in P} \frac{-1}{(z_i - z_j)^2}$ ,

where for a set $N$ $Part_2(N)$ is the set of its perfect matchings.

2) Let $\dim(x) = \dim(y)$. Then

$$per(\tilde{C}(x,y,z)) = per(C(x,y))per\Big(\tilde{C}(z) + Diag\big((-C(z,x) + C(z,y))\vec{1}_{\dim(z)}\big)\Big)$$

Proof:

1) first let's prove that, if for a $d \times d$ -matrix A we define $ham(A) =^{def} = \sum_{\pi \in H_d} \prod_{i=1}^{d} a_{i,\pi(i)}$

where $H_d$ is the set of $d$ -permutations with one cycle, $ham(\tilde{C}(u)) = 0$ when $\dim(u) > 2$.
Suppose $\dim(u) > 2$. Let's partition the set $H_d$ on disjoint subsets consisting of $d$ -1 $d$ -cycles which differ from each other only in the position of the element $d$ . Each of those subsets can be received via taking a ( $d$ -1)-cycle $h = (h_1,...,h_{d-1})$ with elements from the set $\{1,...,d-1\}$ and alternate placing $d$ between neighbors in $h$ (altogether there are $d$ -1 options). In such a case the sum of the corresponding $d$ -1 summands in the expression $ham(\tilde{C}(u))$ will be

$$\sum_{k=1}^{d-1} \frac{u_{h_k} - u_{h_{next(k)}}}{(u_{h_k} - u_d)(u_d - u_{h_{next(k)}})} \prod_{i=1}^{d-1} \frac{1}{u_{h_i} - u_{h_{next(i)}}} = (\prod_{i=1}^{d-1} \frac{1}{u_{h_i} - u_{h_{next(i)}}}) \sum_{k=1}^{d-1} (\frac{1}{u_{h_k} - u_d} - \frac{1}{u_{h_{next(k)}} - u_d}) = 0$$

where $next(t) = \begin{cases} t+1, t < d-1 \\ 1, t = d-1 \end{cases}$ .

Taking into account that for an $n \times n$ -matrix A $per(A) = \sum_{P \in Part(\{1,...,n\})} \prod_{I \in P} ham(A^{(I,I)})$

and $(\tilde{C}(z))^{(I,I)} = \tilde{C}(z_I)$, we get the first claim of the theorem;

2) we apply the induction on $\dim(z)$.
First let's prove the induction's basis for $\dim(z) = 1$. Let $z$ be scalar. Then

$$per(\tilde{C}(x,y,z)) = \lim_{\varepsilon\to 0}(per(C(\begin{pmatrix} x \\ z+\varepsilon \end{pmatrix}, \begin{pmatrix} y \\ z \end{pmatrix})) - \frac{1}{\varepsilon} per(C(x,y))) =$$

$$= \lim_{\varepsilon\to 0}(\frac{\det(C^{\bullet 2}(\begin{pmatrix} x \\ z+\varepsilon \end{pmatrix}, \begin{pmatrix} y \\ z \end{pmatrix}))}{\det(C(\begin{pmatrix} x \\ z+\varepsilon \end{pmatrix}, \begin{pmatrix} y \\ z \end{pmatrix}))} - \frac{1}{\varepsilon}\frac{\det(C^{\bullet 2}(x,y))}{\det(C(x,y))}) =$$

$$= \lim_{\varepsilon\to 0}(\frac{\frac{1}{\varepsilon^2}\det(C^{\bullet 2}(x,y)) + O(1)}{\frac{1}{\varepsilon}\det(C(x,y))(1+\varepsilon(C(z,x)-C(z,y))\vec{1}_{\dim(x)} + O(\varepsilon^2))} - \frac{1}{\varepsilon}\frac{\det(C^{\bullet 2}(x,y))}{\det(C(x,y))})$$

(the second passage is due to the Borchardt formula).

And now the induction's step:

let $\dim(z)=m$, $z_{m+1}$ is a scalar. Then

$$per(\tilde{C}(x,y,\begin{pmatrix} z \\ z_{m+1} \end{pmatrix})) = \lim_{\varepsilon\to 0}(per(\tilde{C}(\begin{pmatrix} x \\ z_{m+1}+\varepsilon \end{pmatrix}, \begin{pmatrix} y \\ z_{m+1} \end{pmatrix}, z)) - \frac{1}{\varepsilon} per(\tilde{C}(x,y,z))) =$$

$$= \lim_{\varepsilon\to 0}\Big(per(C(\begin{pmatrix} x \\ z_{m+1}+\varepsilon \end{pmatrix}, \begin{pmatrix} y \\ z_{m+1} \end{pmatrix}))per(\tilde{C}(z) + Diag((-C(z,\begin{pmatrix} x \\ z_{m+1}+\varepsilon \end{pmatrix}) + C(z,\begin{pmatrix} y \\ z_{m+1} \end{pmatrix}))\vec{1}_{\dim(z)})) -$$

$$-\frac{1}{\varepsilon} per(C(x,y))per(\tilde{C}(z) + Diag((-C(z,x)+C(z,y))\vec{1}_{\dim(z)}))\Big) =$$

$$= \lim_{\varepsilon\to 0}\Big(\frac{1}{\varepsilon} per(C(x,y))(1+\varepsilon(C(z_{m+1},x)-C(z_{m+1},y))\vec{1}_{\dim(x)} + O(\varepsilon^2)) \cdot$$

$$\cdot\big(per(\tilde{C}(z) + Diag((-C(z,x)+C(z,y))\vec{1}_{\dim(z)})) +$$

$$\varepsilon \cdot per(\tilde{C}(\begin{pmatrix} z \\ z_{m+1} \end{pmatrix}) + Diag(\begin{pmatrix} (-C(z,x)+C(z,y))\vec{1}_{\dim(z)} \\ 0 \end{pmatrix})) + O(\varepsilon^2)) -$$

$$-\frac{1}{\varepsilon} per(C(x,y))per(\tilde{C}(z) + Diag((-C(z,x)+C(z,y))\vec{1}_{\dim(z)}))\Big)$$

The last passage is due to the part (1) of this theorem.

## Theorem XI :

$$gen_{\wedge}(u,h,t,\rho) = \lim_{\varepsilon\to 0}(\varepsilon^{2\dim(u)} gen_{\sim}(\begin{pmatrix} u \\ u+\sqrt{1+\sqrt{-1}\varepsilon}\vec{1}_{\dim(u)} \\ u+\sqrt{1-\sqrt{-1}\varepsilon}\vec{1}_{\dim(u)} \end{pmatrix}, \begin{pmatrix} \vec{0}_{\dim(u)} \\ \vec{1}_{\dim(u)} \\ -\vec{1}_{\dim(u)} \end{pmatrix}, h,t,\rho))$$

Proof:

$$\lim_{\varepsilon\to 0}(\varepsilon^{2\dim(u)} gen_{\sim}(\begin{pmatrix} u \\ u+\sqrt{1+\sqrt{-1}\varepsilon}\vec{1}_{\dim(u)} \\ u+\sqrt{1-\sqrt{-1}\varepsilon}\vec{1}_{\dim(u)} \end{pmatrix}, \begin{pmatrix} \vec{0}_{\dim(u)} \\ \vec{1}_{\dim(u)} \\ -\vec{1}_{\dim(u)} \end{pmatrix}, h,t,\rho)) = gen_{\sim}(u, \sqrt{-1}\vec{1}_{\dim(u)}, h,t,\rho) =$$

$$= \sum_{I} \sum_{J,|J|=\dim(h)} (\sqrt{-1})^{|I|} per(\tilde{C}(u_I)) per(\tilde{C}(h,t_J,u_{\setminus I})) \prod_{j\in J} \rho_J =$$

(we use Theorem X(2) )

$$= \sum_{I} \sum_{J,|J|=\dim(h)} (\sqrt{-1})^{|I|} per(\tilde{C}(u_I)) per(\tilde{C}(u_{\setminus I}) + Diag((-C(u_{\setminus I},h) + C(u_{\setminus I},t_J))\vec{1}_{\dim(u)})) per(C(h,t_J)) \prod_{j\in J} \rho_J =$$

(we use Theorem X(1) )

$$= \sum_{J,|J|=\dim(h)} per(Diag((-C(u,h)+C(u,t_J))\vec{1}_{\dim(u)})) per(C(h,t_J)) \prod_{j\in J} \rho_J =$$

$$= \sum_{J,|J|=\dim(h)} (\prod_{i=1}^{\dim(u)} (-C(u_i,h)\vec{1}_{\dim(u)} + C(u_i,t_J)\vec{1}_{\dim(u)})) per(C(h,t_J)) \prod_{j\in J} \rho_J = gen_{\wedge}(u,h,t,\rho)$$

## Theorem XII :

Let $\dim(u)=\dim(h), C^{\bullet(\frac{1}{3},1)}(h,t)Diag(\rho)=\vec{0}_{2\dim(h)}$. Then

$$per(C^{\bullet 3^m}(u,t)Diag(\rho)C(t,h)) = (1+\sqrt{-1})^{\dim(h)} gen_{\wedge}(u\otimes\vec{1}_{3^m}, h, \begin{pmatrix} t \\ t \end{pmatrix}, \begin{pmatrix} \rho \\ \rho\sqrt{-1} \end{pmatrix})$$

Proof:

It follows from the generalized Binet-Minc identity for $p\hat{e}r_B(A)$, due to the fact that $\forall I, |I|\le 3, scal\big((C(h,\begin{pmatrix} t \\ t \end{pmatrix})Diag(\begin{pmatrix} \rho \\ \rho\sqrt{-1} \end{pmatrix}))^{(I,\{1,\dots,2\dim(t)\})}\big)=0$ (by the construction and the condition $C^{\bullet(\frac{1}{3},1)}(h,t)Diag(\rho)=\vec{0}_{2\dim(h)}$), while $\dim(u)=\dim(h)$.

## Corollary XII.1 :

Let $\dim(u)=\dim(h)=n=3^m$ , $\dim(t)=n(n+2), \det(\begin{pmatrix} C^{\bullet(\frac{1}{3},1)}(h,t) \\ C^{\bullet(1,\dots,n)}(u,t) \end{pmatrix})\neq 0,$

$M$ is an $n\times n$-matrix. Then

$$per(M) = (1+\sqrt{-1})^{-n} gen_{\wedge}(u\otimes\vec{1}_{3^m}, h, \begin{pmatrix} t \\ t \end{pmatrix}, (\begin{pmatrix} C^{\bullet(\frac{1}{3},1)}(h,t) \\ (C^{\bullet(1,\dots,n)}(u_1,h))^T C^{\bullet(n,\dots,1)}(u_1,t) \\ \dots\dots\dots \\ (C^{\bullet(1,\dots,n)}(u_n,h))^T C^{\bullet(n,\dots,1)}(u_n,t) \end{pmatrix}^{-1} \begin{pmatrix} \vec{0}_{2n} \\ M^{(\{1,\dots,n\},1)} \\ \dots \\ M^{(\{1,\dots,n\},n)} \end{pmatrix}) \otimes \begin{pmatrix} 1 \\ \sqrt{-1} \end{pmatrix})$$